\documentclass[sigconf]{acmart}

\usepackage{amsmath,amsfonts,bm}

\def\eqref#1{equation~\ref{#1}}

\def\1{\bm{1}}

\DeclareMathAlphabet{\mathsfit}{\encodingdefault}{\sfdefault}{m}{sl}
\SetMathAlphabet{\mathsfit}{bold}{\encodingdefault}{\sfdefault}{bx}{n}

\usepackage[ruled,vlined]{algorithm2e}
\usepackage{color}
\usepackage{colortbl}
\usepackage{import}
\usepackage{paralist}
\usepackage{subcaption}
\usepackage{amsmath} 
\usepackage{caption}
\usepackage{multirow}
\usepackage{makecell}
\usepackage{threeparttable}
\usepackage{graphicx} 
\usepackage{enumitem}
\setlist[itemize]{leftmargin=*}

\AtBeginDocument{%
  \providecommand\BibTeX{{%
    \normalfont B\kern-0.5em{\scshape i\kern-0.25em b}\kern-0.8em\TeX}}}

\setcopyright{acmcopyright}
\copyrightyear{2018}
\acmYear{2018}
\acmDOI{XXXXXXX.XXXXXXX}

\acmConference[XXX]{XXXX}{June 03--05, 2021}{XXXX}
\acmBooktitle{WXXXX,
  June 03--05, 2021, XXX, XX}
\acmPrice{15.00}
\acmISBN{978-1-4503-XXXX-X/18/06}

\graphicspath{./img}

\begin{document}


\title{Data-Augmented Counterfactual Learning for Bundle Recommendation}


\author{Shixuan Zhu}
\affiliation{%
  \institution{Tongji University}
  \city{Shanghai}
  \country{China}}
\email{2130768@tongji.edu.cn}

\author{Qi Shen}
\affiliation{%
  \institution{Tongji University}
  \city{Shanghai}
  \country{China}}
\email{1653282@tongji.edu.cn}

\author{Yiming Zhang}
\affiliation{%
  \institution{Tongji University}
  \city{Shanghai}
  \country{China}}
\email{2030796@tongji.edu.cn}

\author{Zhenwei Dong}
\affiliation{%
  \institution{Tongji University}
  \city{Shanghai}
  \country{China}}
\email{1853155@tongji.edu.cn}

\author{Zhihua Wei}
\authornote{Corresponding author.}
\affiliation{%
  \institution{Tongji University}
  \city{Shanghai}
  \country{China}}
\email{zhihua_wei@tongji.edu.cn}

\renewcommand{\shortauthors}{XXX, et al.}


\begin{abstract}
Bundle Recommendation (BR) aims at recommending bundled items on online content or e-commerce platform, such as song lists on a music platform or book lists on a reading website.
Several graph-based models have achieved state-of-the-art performance on BR task.
But their performance is still sub-optimal, since the data sparsity problem tends to be more severe in real bundle recommendation scenarios,
which limits graph-based models from more  sufficient learning. 
In this paper, we propose a novel graph learning paradigm called Counterfactual Learning for Bundle Recommendation (CLBR) to mitigate the impact of data sparsity problem and improve bundle recommendation. 
Our paradigm consists of two main parts: counterfactual data augmentation and counterfactual constraint.
The main idea of our paradigm lies in answering the counterfactual questions: "What would a user interact with if his/her interaction history changes?" "What would a user interact with if the
bundle-item affiliation relations change?"
In counterfactual data augmentation, we design a heuristic sampler to generate counterfactual graph views for graph-based models, which has better noise controlling than the stochastic sampler.
We further propose counterfactual loss to constrain model learning for mitigating the effects of residual noise in augmented data and achieving more sufficient model optimization.
Further theoretical analysis demonstrates the rationality of our design.
Extensive experiments of BR models applied with our paradigm on two real-world datasets are conducted to verify the effectiveness of the paradigm.
\end{abstract}

\ccsdesc[500]{Information systems~Recommender systems}

\keywords{Bundle Recommendation, Data Augmentation, Counterfactual Learning}


\maketitle


%
%
\vspace{-0.2cm}
\section{Introduction}
\vspace{-0.1cm}

Bundled items are very common on nowadays online content or e-commerce platforms, which makes Bundle Recommendation (BR) \cite{zhu2014bundle} become an important task.
The task aims to recommend a bundle with multiple items for users, which can improve users' experience for providing more variable options, as well as increase business profits for the expansion of order size.
Typical BR scenarios include book lists recommendation on online reading websites \cite{liu2014list}, song lists recommendation on music streaming platforms, video collections recommendation on online video websites, etc.


On the basis of general recommendation methods, most existing works for BR introduce user-item interaction and bundle-item affiliation information as supplements for user-bundle interaction.
For utilizing these key information, graph-based models are employed\cite{GCN,rgcn} to encode them into user and bundle embeddings \cite{BGCN,BundleNet,CrossCBR}, and some multi-task frameworks \cite{cao2017embedding,DAM,BundleNet} are leveraged for model training.

However, the effectiveness of the aforementioned graph-based BR models usually depends on enough high-quality training data.
Unfortunately, the data sparsity problem is more severe and complex in most realistic BR scenarios than in general recommendation, since there exists not only user-side sparsity but also creator-side sparsity problems.
For the users of platform, similar to the ubiquitous user-item interaction sparsity \cite{Datasparsity,causerec,improving}, the user-bundle interactions are also very sparse.
For the bundle creators, although there are massive possible item combinations for bundle generation, very few bundles are actually created and shared \cite{BuildBundle}, which leads to sparse bundle-item affiliations.
These prevent adequate training for the graph-based BR models, and lead to sub-optimal recommendation results.
%

For confronting these data sparsity in BR, we focus on the counterfactual thinking in causal theory and resort to counterfactual data as a complement to the observed sparse data.
In counterfactual thinking, the real data we observe in BR scenarios, whether interaction data or bundle-item affiliation relation, is only a small fraction of all possible data, or counterfactual data.
\begin{figure}[t]
\vspace{-0.2cm}
    \centering
    \includegraphics[width=0.45\textwidth]{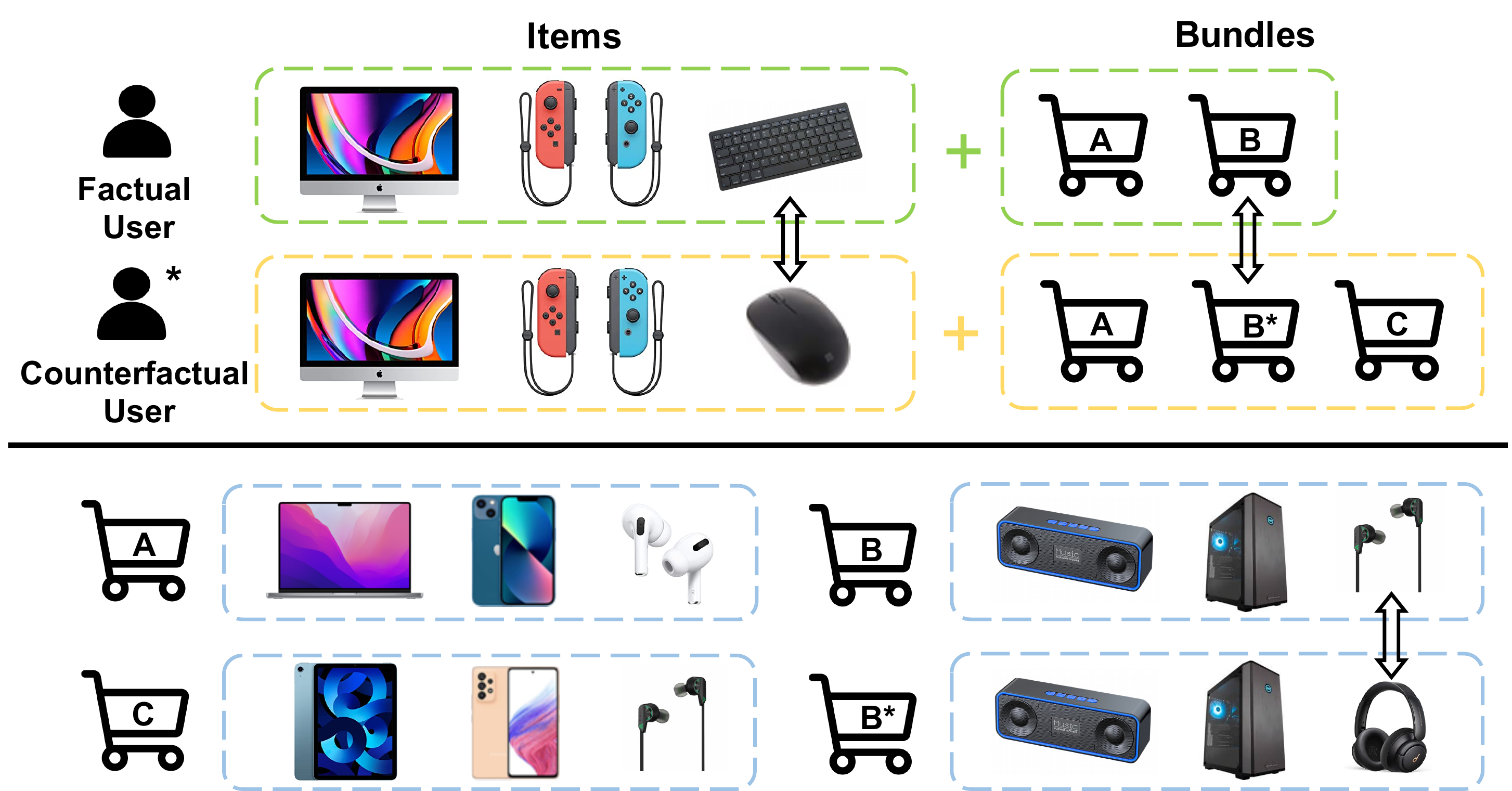}
    \caption{A toy example of bundle recommendation in E-commerce, where the bundles are shopping carts that  consist of multiple products.}
    \label{fig:toy}
    \vspace{-0.1cm}
\end{figure}
For instance, Figure \ref{fig:toy} illustrates a toy example of BR in E-commerce, where a user can explore not only products, but also shopping carts that consist of multiple products created by other users.
We observe that a user purchased a keyboard instead of a mouse, and only chose \emph{bundle A} and \emph{bundle B}.
But in the counterfactual world, due to possible different exposure mechanisms, the user may have bought the mouse, and additionally selected \emph{bundle C} that has similar contents with \emph{bundle A}.
Or perhaps the contents of \emph{bundle B} changes due to the substitutability of some items, but the user still can accept the changed bundle.
These counterfactual data can more comprehensively reveal users' preference, which naturally solve the data sparsity problem and contribute to more effective model training.

Motivated by the analysis above, we start with the following "what if" questions:
\textbf{"What would a user interact with if his/her interaction history changes?"}
\textbf{"What would a user interact with if the bundle-item affiliation relations change?"} and try to simulate these counterfactuals by intervening on the user's interaction history and bundle-item affiliation for improving BR model training.
We propose a novel learning paradigm called Counterfactual Learning for Bundle Recommendation (CLBR) to generate counterfactual data and guide the graph-based BR model to utilize these data effectively.
It consists of two main parts: counterfactual data augmentation and counterfactual constraint. 
For the data augmentation, we develop a simple yet efficient heuristic sampler to generate counterfactual graph views based on the original BR graph, which achieves effective noise control compared to the stochastic sampler. 
The counterfactual constraint helps models discard intractable noise in the generated data that deviates from the counterfactual distribution, and make efficient use of augmented data to learn accurate and robust representations for BR. 
We apply the CLBR paradigm to several state-of-the-art graph-based models. 
Extensive experiments on two real-world datasets demonstrate the effectiveness of the proposed paradigm.

Our main contributions in this work are summarized below:
\begin{itemize}[leftmargin=*]
    \item We introduce the data sparsity problem in bundle recommendation, and try to leverage the counterfactual thinking into bundle recommendation for confronting the problem.
    \item We propose \textbf{C}ounterfactual \textbf{L}earning for \textbf{B}undle \textbf{R}ecommendation (CLBR), a novel learning paradigm for graph-based models, which includes counterfactual data augmentation with a heuristic sampler and counterfactual constraint for training, to achieve more sufficient model optimization.
    \item We theoretically analyze the noisy information in the augmented data, and interpret our proposed constraint term via information theory, to verify the rationality of our design.
    \item We apply CLBR on several state-of-the-art graph-based models and conduct extensive experiments based on real-world BR datasets, which verify the effectiveness of the paradigm.
\end{itemize}


\vspace{-0.1cm}
\section{Related Works}
\vspace{-0.05cm}
\textbf{Bundle Recommendation.}
Recently, several efforts have been made in Bundle Recommendation, which recommends a bundle of items that users may have interest to interact together\cite{sar2016beyondCF,liu2017modeling,EFM,DAM,BGCN,HGCN,CrossCBR}.
Sar Shalom et al. \cite{sar2016beyondCF} introduced the list recommendation problem, and optimized a list's click probability based on collaborative filtering.
Liu et al. \cite{liu2017modeling} estimated the probability that a consumer would buy an item associated with already bought item to recommend product bundles.
EFM \cite{EFM} captured users’ preferences over items and lists by factorization model.
DAM \cite{DAM} contributed a factorized attention network to gather item information into bundle representations in a multi-task way jointly for user-bundle and user-item interactions.
More recently, graph-based recommendation models have become the state-of-the-art methods for BR.
HFGN \cite{HGCN} proposed a Hierarchical Fashion Graph Neural Network to obtain more expressive representations for users and outfits, which incorporated both item-level and bundle-level semantics into the bundle representations.
BGCN \cite{BGCN} utilized graph neural network (GNN) to learn representations from  user-level and bundle-level graph views, and CrossCBR\cite{CrossCBR} further adopted contrastive learning to model the cooperative association between these two views to achieve mutual enhancement. 

However, previous researches rely on large amount of reliable data.
The severe data sparsity problems existed in BR scenarios limit the performance of graph-based models, which inspires the study of counterfactual data augmentation in this paper.

\noindent \textbf{Counterfactual View in Recommendation.}
As a key theory of causal inference, counterfactual paradigm focuses on estimating causal effects with \emph{What if} problems and has been widely adopted in many machine learning domains recently, including computer vision \cite{CVE,CVL}, natural language processing \cite{CNLP} and graph mining \cite{sui2021CGNN,C_GNN_link}.
In the recommendation field, some works leveraged counterfactual thinking for debias problem.
For example, Wang et al. \cite{C_RS_clicks} estimated the counterfactual click likelihood of the user for reducing the direct effect of exposure features and eliminating the clickbait issue.
Zhang et al. \cite{C_RS_Popularity} deconfounded the popularity bias in training phase, and adjusted the recommendation score with desired popularity information via causal intervention during inference phase. 
Besides, some researchers improved the recommendation models in causal view.
Zhang et al. \cite{zhang2021causalACF} empowered attention mechanism by the causal regularization, which uses individual treatment effect to measure the causal relation.
In addition, some methods explored the counterfactual in data-augmentation manner \cite{xu2021CCF}.
For instance, Wang et al. \cite{CasualSR} proposed data- and model-oriented sampling methods for generating better counterfactual sequences when training the anchor model.
Xiong et al. \cite{xiong2021Creview} designed a learning-based method to discover more effective samples, actively intervened on the user preference, and predicted the user feedback based on a pre-trained recommender.

However, the counterfactual view in the heterogeneous graph field, e.g. the complex multi-relation in BR, is of great need but largely unexplored. 

\vspace{-0.1cm}
\section{Preliminary}\label{sec:define}
\vspace{-0.05cm}

\subsection{Task Definition}
\vspace{-0.05cm}
Given user set $U = \{u_1, u_2, \dots, u_N\}$, item set $V = \{v_1, v_2,\dots, v_L\}$, and bundle set $B = \{b_1, b_2, \dots, b_K\}$, we formulate the user-item interaction data as $\mathcal{E}_{uv}=\{(u,v)| u\;interacted\; v,u\in U, v\in V\}$, user-bundle interaction data as $\mathcal{E}_{ub}=\{(u,b)| u\;interacted\; b,u\in U, b\in B\}$ and bundle-item affiliation relation as $\mathcal{E}_{bv}=\{(b,v)\mid v\;belongs\; to\; b,b\in B, v\in V\}$.
Based on the above information, bundle recommender estimates the probability that user will interact with each candidate bundle.

\vspace{-0.25cm}
\subsection{A Unified Framework}
\vspace{-0.05cm}
Then we summarize previous bundle recommendation methods into a unified graph neural network framework for subsequent leveraging. 
Generally, we organize the user, item and bundle and their relation information as a heterogeneous user-bundle-item graph $\mathbf{G}=\{\mathcal{V}=\{U,V,B\},\mathcal{E}=\{\mathcal{E}_{uv},\mathcal{E}_{ub},\mathcal{E}_{bv}\}\}$ following \cite{BGCN,HGCN}, with initialized node embeddings $\mathbf{X}\in \mathbb{R}^{(N+L+K)\times D}$ of $\mathcal{V}$, where $D$ is the embedding dimension. 
Then we use the unified user-bundle-item relation graph defined above as the basis of graph-based BR models.
The general structure of existing graph-based BR models can be summarized in the following two parts:
(1) modeling the cross-semantic relation of user, item and bundle; (2) calculating the final matching score of user and bundle.
Therefore, we formula the unified bundle recommender $M_{\theta}(\mathbf{X},\mathbf{G})$ mainly as a graph-based encoder and an embedding-based predictor: 
\begin{equation}
\setlength{\abovedisplayskip}{3pt}
\setlength{\belowdisplayskip}{3pt}
\begin{aligned}
\mathbf{U},\mathbf{B},\mathbf{I}=\mathbf{Encoder}(\mathbf{X},\mathbf{G}) \;;\;\;
s_{i,j}=\mathbf{Predictor}(\mathbf{U}_i,\mathbf{B}_j)
\end{aligned}
\label{eqn:predictor}
\end{equation}
where $\mathbf{U},\mathbf{B},\mathbf{I}$ denote the learned embeddings of users, bundles and items respectively, and $s_{i,j}$ is the predicted probability of user $i$ and bundle $j$ based on their embeddings.
The loss objective $L_{tasks}(M_{\theta})$ can be designed flexibly for various operations, e.g. contrastive learning \cite{CrossCBR}, hard negative sampling \cite{BGCN}.
Through the minimization of loss, the bundle recommender parameters are optimized.






\vspace{-0.15cm}
\section{Methodology}\label{sec:model}
\vspace{-0.05cm}

\begin{algorithm}
  \SetAlgoLined
  \KwData{A unified user-bundle-item relation graph $\mathbf{G}^f$ ; Number of counterfactual views $N_{v}$; Initialized embeddings $\mathbf{X}$ of users, items and bundles.}
  \KwResult{Learned bundle recommendation model $M_\theta$.}
  Pretrain the selection model $M_\theta'$\;
  \For(\tcp*[f]{Counterfactual Data Aug}){i in $[1,N_{v}]$}{  
    \For{$t$ in $[ub,ui,bi]$}{
    $\mathcal{E}_{t}^{+}= \emptyset, \quad \mathcal{E}_{t}^{-}= \emptyset  $\;
    \While{$|\mathcal{E}_{t}^{+}|+|\mathcal{E}_{t}^{-}|\le r_{t}|\mathcal{E}_{t}|$}{
    Sample a batch of node pairs under $t$, and add them into $\mathcal{E}_{t}^+$ and $\mathcal{E}_{t}^-$ following Eq. \ref{eqn:generate_rule}\; 
        }
    }
    Append the $\mathbf{G}^{c}=\{\mathcal{V},\mathcal{E}^c\}$ into counterfactual view set $\mathcal{G}^{c}$, where $\mathcal{E}^c=\mathcal{E} \cup \sum_t \mathcal{E}^+_t -\sum_t \mathcal{E}^-_t$\; 
  }
  \For(\tcp*[f]{Counterfactual Constraint}){epoch in $[0,T]$}{
    Randomly choose a counterfactual view $\mathbf{G}^{c} \in \mathcal{G}^{c}$\;
    Generate the user and bundle representations via bundle recommendation model:
    $\mathbf{U}^f,\mathbf{B}^f=M_{\theta}(\mathbf{X},\mathbf{G}^f)$, $\mathbf{U}^{c},\mathbf{B}^{c}=M_{\theta}(\mathbf{X},\mathbf{G}^{c})$ \;
    Normalize the embeddings of user and bundle   \;
    Calculate the loss $\mathcal{L}$ through Eq. \ref{eqn:loss_total}\;
    Update $\theta$ by gradient descent\;
    }
    \textbf{Inference:} Calculate the interaction probability score through $M_{\theta}(\mathbf{X},\mathbf{G}^f)$ with frozen parameters $\theta$. 
  \caption{Learning Algorithm of CLBR}    
  \label{alg:ccbr}
\end{algorithm}
\vspace{-0.25cm}

In this section, we introduce the proposed counterfactual learning paradigm to alleviate the problem of sparsity effectively in BR scenario.
The complete process of our proposed Counterfactual Learning for Bundle Recommendation (CLBR) is presented in Algorithm \ref{alg:ccbr}, which consists of two main steps: counterfactual data augmentation with a heuristic node pairs sampler, and model training under counterfactual constraint.
\subsection{Counterfactual Data Augmentation}\label{sec:datagen}
In this part, we generate a set of counterfactual views based on the unified user-bundle-item relation graph as data augmentation.
In detail, we use a sampler $S$ to choose user-bundle, user-item and bundle-item pairs, and then change the structure of the original user-bundle-item relation graph by adding/dropping edges according to the sampling results.
We set the augmentation ratios $r_{ui},r_{ub},r_{bi}$ for three types of node pairs respectively, to control the disturbance ratio of counterfactual views.
We denote that $\mathcal{E}^+_t$ is the set of node pairs selected to be added as edges to the counterfactual graph view, while $\mathcal{E}^-_t$ is the set of edges picked for dropping, where $t$ denotes the type of node pairs which can be $ui$,$ub$ and $bi$.

There are many feasible methods to implement the sampler.
In the simplest stochastic sampling method, we randomly select node pairs for $\mathcal{E}^+_t$ from all unconnected node pairs in the original BR graph, as well as select node pairs for $\mathcal{E}^-_t$ from existing edges in BR graph, until the size of $\mathcal{E}^+_t$ and  $\mathcal{E}^-_t$ for each type satisfy the ratio we set. 
Formally, it can be expressed as
$|\mathcal{E}^{+}_{t}|\ge \alpha r_{t}|\mathcal{E}_{t}|, |\mathcal{E}^{-}_{t}|\ge (1-\alpha) r_{t}|\mathcal{E}_{t}|$ for $\forall t$, where $\alpha \in [0,1]$ controls the proportion of adding and dropping.
Then we can generate a counterfactual view of BR graph by applying these user-bundle, user-item and bundle-item edge adding and dropping operations to the original graph.
Finally, we get a counterfactual view set through generating $N_{v}$ counterfactual views, which is a naive simulation of the counterfactual world.

The stochastic mechanism is easy to implement, but it brings too much noise to the generated graph views and reduces the effectiveness of subsequent model learning.
So we designed a simple yet efficient heuristic sampling method in Algorithm \ref{alg:ccbr}, which improve the reliability of sampling by considering noise control.
We pre-train a selection model $M_{\theta'}$ which has the same structure of the recommendation model $M_{\theta}$, and utilize its graph encoder to generate all user, bundle and item embeddings $U,B,I$ before we start the sampling phase.
Considering the computational complexity, we randomly sample a batch of node pairs each time, query the embeddings of them, and calculate relevance scores for the node pairs in the batch.
Taking user-bundle node pair $(i,j)$ as an example, the relevance scores can be calculated as $rs_{i,j} = {U_i^\top}B_j$.
Then we filter the noisy data by the following selection rule: 
\begin{equation}
\setlength{\abovedisplayskip}{3pt}
\setlength{\belowdisplayskip}{3pt}
\begin{aligned}
\left\{\begin{aligned} &\text{$\mathcal{E}^+_{ub}$} \leftarrow \text{$\mathcal{E}^+_{ub}$}\cup\{\left(i, j\right)\},\;\;\;if \; rs_{i,j}  > {\kappa}^+ and \left(i, j\right) \not\in \mathcal{E}_{ub} 
\\ &\text{$\mathcal{E}^-_{ub}$} \leftarrow \text{$\mathcal{E}^-_{ub}$}\cup\{\left(i, j\right)\},\;\;\;if\;  rs_{i,j} \leq \kappa^- and \left(i, j\right) \in \mathcal{E}_{ub} 
\\ &\text{No selection,} \;\;\;\; {else} \end{aligned}\right.
\end{aligned}
\label{eqn:generate_rule}
\end{equation}
where ${\kappa}^+$ and ${\kappa}^-$ are batch-based thresholds for selection, which are defined as the maximum and minimum relevance scores of user-bundle pairs in the batch multiplied by fixed ratios $\alpha^+ $ and $\alpha^-$ respectively.
This design simplifies the complex computation for full collection node pairs to the computation within every batch while ensuring noise filtering for effective augmentation.
By repeating randomly sampling and selection process, we keep filling the adding set $\mathcal{E}^+_t$ and dropping set $\mathcal{E}^-_t$ until they contain  user-bundle, user-item and bundle-item node pairs with total numbers of $r_{ub}|\mathcal{E}_{ub}|$, $r_{ui}|\mathcal{E}_{ui}|$ and $r_{bi}|\mathcal{E}_{bi}|$ respectively. 
Then we merge $\mathcal{E}^+_t$ into $\mathcal{E}$ and remove $\mathcal{E}^-_t$ from $\mathcal{E}$ for all $t$ to generate a counterfactual view, and further get a counterfactual view set by generating $N_{v}$ counterfactual views. 

\vspace{-0.05cm}
\subsection{Counterfactual Constraint}\label{sec:phase3}
Then the question we would like to answer is: given the augmented data composed of the original factual graph and generated counterfactual views, how to achieve effective learning for BR model $M_\theta$?
Despite our noise control in the data augmentation process, some noise that deviates from the counterfactual distribution will inevitably remain in the generated data.
In order to minimize the influence of residual noise, the model training needs to be additional constrained.
Although the interaction data in the counterfactual world is different from that in the real world, user preferences, bundle and item features should be stable. 
While using GNN to perform semantic propagation on the counterfactual view and the original graph respectively, we expect that the embeddings of the same node learned on different graph views should be closer, while the distance between the node embedding distributions of the two views should vary within a reasonable range.

Therefore, for the normalized embeddings of user $\mathbf{U}$ and bundle $\mathbf{B}$, we have the following counterfactual constraints:
\begin{equation}
\setlength{\abovedisplayskip}{3pt}
\setlength{\belowdisplayskip}{3pt}
\begin{aligned}
    & \text{minimize} \quad  L_{tasks}(M_{\theta}) 
    \\&\
    \text{s.t.} \sum_{i=1}^N \mathcal{D}(\mathbf{U}^{c}_i,\mathbf{U}^{f}_i) \leq \lambda_u \sum_{i=1}^N \sum_{j=1,j \neq i}^N \mathcal{D}(\mathbf{U}^{c}_j,\mathbf{U}^{f}_i), 
    \\&\
    \;\;\;\;    \sum_{i=1}^K \mathcal{D}(\mathbf{B}^{c}_i,\mathbf{B}^{f}_i) \leq \lambda_b \sum_{i=1}^N \sum_{j=1, j\neq i}^K\mathcal{D}(\mathbf{B}^{c}_j,\mathbf{B}^{f}_i) ,\forall c \in \mathcal{G}^c
\end{aligned}
\label{eqn:loss_org}
\end{equation}
where the function $\mathcal{D}(e_i,e_j)$ represents the distance between embeddings of node pair $e_i$ and $e_j$ in latent space, $c$ denotes one potential counterfactual augmentation.
The left consistency term in the two inequality reflects the consistency between the embeddings in the original graph and counterfactual view of the same user and bundle separately, and the right unrestraint term reflects the \emph{average consistency} of every single node's embedding on the original graph with the embeddings of all other nodes in the augmented counterfactual view.
Compared to the constant threshold $\lambda_u,\,  \lambda_b$ for the consistency between the real relation and counterfactual relation \cite{xu2021CCF}, we adopt the additional \emph{average consistency} metrics for self-adaptive threshold settings.
With lower \emph{average consistency}, which can be explained as the user or bundle node being further from other nodes in latent space, we can make larger edge-level perturbations while ensuring that the node will not be confused with other nodes due to changes in embeddings caused by the perturbations.
Through the above two constraints, the appropriate-intervention information is preserved and over-intervention information is discarded for exclusive user and bundle representations.

Directly solving the above constrained optimization problem is challenging for the constraints are not differentiable. 
Therefore we relax the constraints and convert the objective in Eq. \ref{eqn:loss_org} to the Lagrange optimization form as follows:
\begin{equation}
\setlength{\abovedisplayskip}{2pt}
\setlength{\belowdisplayskip}{2pt}
\begin{aligned}
\text{minimize} \quad L=L_{tasks}(M_{\theta}) + \omega_u L_u + \omega_b L_b
\end{aligned}
\label{eqn:loss_total}
\end{equation}
where non-negative hyper-parameter $\omega_u,\, \omega_b$ control the weight of the user- and bundle-side constraint loss $ L_u,\, L_b$ respectively, which are defined as follows:
\begin{equation}
\setlength{\abovedisplayskip}{2pt}
\setlength{\belowdisplayskip}{0pt}
\begin{aligned}
L_u = \frac{1}{N}\sum_{i=1}^N \left(\mathcal{D}(\mathbf{U}^{c}_i,\mathbf{U}^{f}_i) - \lambda_u\sum_{j=1,j \neq i}^N \mathcal{D}(\mathbf{U}^{c}_j,\mathbf{U}^{f}_i)\right)
\end{aligned}
\label{eqn:loss_lu}
\end{equation}
\begin{equation}
\setlength{\abovedisplayskip}{2pt}
\setlength{\belowdisplayskip}{2pt}
\begin{aligned}
L_b = \frac{1}{K}\sum_{i=1}^K \left( \mathcal{D}(\mathbf{B}^{c}_i,\mathbf{B}^{f}_i) - \lambda_b\sum_{j=1, j\neq i}^K\mathcal{D}(\mathbf{B}^{c}_j,\mathbf{B}^{f}_i) \right)
\end{aligned}
\label{eqn:loss_lb}
\end{equation}
In our experiments, the distance metric $\mathcal{D}(e_i,e_j)$ is computed as $-\exp(e_i^Te_j/\tau)$, where $\tau$ denotes the temperature hyper-parameter for tuning how concentrated the features are in the latent space.
In practice, $ \mathcal{D}(e_i,e_j)$ can be any other distance metric function.
To reduce computational complexity, the average global representations are not explicitly sampled but generated from the other user/bundle nodes within the same minibatch.
And the final loss is computed across all representations of users/bundles in the minibatch.



\vspace{-0.15cm}
\section{Further Discussion}\label{sec:think}
\vspace{-0.05cm}

\subsection{Analysis about Noisy Control}\label{sec:data-aug}
The data augmentation process in CLBR depends on sampler $S$. 
If $S$ is not accurate, the augmented data can be noisy.
To check the necessity of noise control in counterfactual data augmentation process, we would like to know how many data should be generated when given the noise level, if we want to ensure relatively good performance. 
So we theoretically analyze the relation between the amount of generated data and the noisy information introduced by $S$, under the pursuit of well performance based on PAC learning framework\cite{PAC2014}. 
We assume that our sampler can recover the true edges in the counterfactual graph views of the user-item-bundle graph with the probability of $1-\eta$, where $\eta \in(0,0.5)$ indicates the noisy level of $S$.
Then we have the following theory:

\begin{theorem}{}
Given a hypothesis class $\mathcal{H}$, for any $\epsilon, \delta \in(0,1)$ and $\eta \in(0,0.5)$, if $h \in \mathcal{H}$ is the edge ranking model learned based on the empirical risk minimization (ERM), and the sample complexity (i.e., number of samples) is larger than $\frac{2 \log \left(\frac{2|\mathcal{H}|}{\delta}\right)}{\epsilon^{2}(1-2 \eta)^{2}}$, then the error between the model estimated and true results is smaller than $\epsilon$ with probability larger than $1-\delta$.
\end{theorem}

The proof of this theorem can refer to \cite{CasualSR}.
Assuming that the generated data is noisy, if the prediction error of $h$ is larger than $\epsilon$, we have the empirical mismatching rate of $h$ is smaller than $\eta+\frac{\epsilon(1-2 \eta)}{2}$, or the empirical mismatching rate of the optimal $h^*$ is larger than $\eta+\frac{\epsilon(1-2 \eta)}{2}$.
In order to achieve a good performance with given probability (i.e., $\epsilon$ and $\delta$), at least $\frac{2 \log \left(\frac{2|\mathcal{H}|}{\delta}\right)}{\epsilon^{2}(1-2 \eta)^{2}}$ samples are needed.
That is to say, when the noise level of sampler $\eta$ is larger, we need to choose much more edges for generating the counterfactual views, which is a huge computation consumption.
This theory reveals the significance of our designed heuristic sampler with noise control.
The adjustable threshold for selection rule guarantees a trade-off between sample efficiency and noise level, so the heuristic sampler can generate counterfactual graph views with better reliability and outshone the stochastic one in the case of generating the same amount of data.

\subsection{An Information Theory Interpretation}\label{sec:GCL}
In this section, we provide some analysis of the proposed counterfactual loss design.
Here, take the Eq. \ref{eqn:loss_lu} as an example, we rewrite it as $L_{u}= L_{consis} + \lambda_{u} L_{unres}$ and analysis the consistency term $L_{consis}= \frac{1}{N} \sum_{i}\mathcal{D}\left(\mathbf{U}^{c}_i,\mathbf{U}^{f}_i\right)$ and unrestraint terms $L_{unres}= \frac{-1}{N}\sum_{i} \sum_{i\neq j}{\mathcal{D}\left(\mathbf{U}^{c}_i,\mathbf{U}^{f}_j\right)}$ respectively.
Here, $I(\cdot,\cdot)$ is used to measure the mutual information between two variables, $I(\cdot|\cdot)$ for conditional mutual information, $H(\cdot)$ for the entropy, and $H(\cdot|\cdot)$ for conditional entropy. 
Then, under a general assumption that the distributions of $P(\mathbf{U}^{c})$ and $P(\mathbf{U}^{c}|\mathbf{X},\mathbf{G})$ obeys a Gaussian distribution, we can arrive the following two lemmas:
\begin{lemma}
Minimizing $L_{consis}$ is equivalent to minimizing the entropy of $\;\mathbf{U}^{c}$ conditioned on input data $\mathbf{X}$ and $\mathbf{G}$, i.e., $\underset{\theta}{\min}\;L_{consis}  \cong \underset{\theta}{\min}\;H( \mathbf{U}^{c} |\mathbf{X}, \mathbf{G}) $.
\end{lemma}
\vspace{-0.2cm}
\begin{proof}
For the factual view $f$ and counterfactual view $c$ of $\mathbf{G}$ through counterfactual augmentation, i.e. $f,c\sim p(\cdot|\mathbf{G})$, we have:
\begin{equation}
\setlength{\abovedisplayskip}{2pt}
\setlength{\belowdisplayskip}{2pt}
\begin{aligned}
& L_{consis} \propto -2\sum_{i=1}^N {\sum_{j=1}^D{}} {\mathbb{E} _{f,c \sim p\left( \cdot |\mathbf{G} \right)}} \mathbf{U}^{c}_{ij}\mathbf{U}^{f}_{ij} + 2N
 \\&\
=\sum_{i=1}^N{\sum_{j=1}^D{}} {\mathbb{E} _{f,c\sim p\left( \cdot |\mathbf{G} \right)}}\left( \mathbf{U}^{c\;2}_{ij}+\mathbf{U}^{f\;2}_{ij}-2\mathbf{U}^{c}_{ij}\mathbf{u}^{f}_{ij} \right) 
 \\&\
=2\sum_{i=1}^N{\sum_{j=1}^D{}} \mathbb{D} _{v \sim p\left( \cdot |\mathbf{G} \right)}\mathbf{U}^{v}_{ij} 
=2\sum_{i=1}^N{\mathbb{D} _{v \sim p\left( \cdot |\mathbf{G} \right)}\mathbf{U}^{v}_i }
\end{aligned}
\label{eqn:proof1}
\end{equation}
where $v$ represents the  view of $f$ and $c$.
This indicates that minimizing $L_{consis}$ is to minimize the variance of counterfactual and factual’s representations conditioned on the input data.
Note that the unrestraint term $L_{unres}$ aims to make different representations be roughly uniformly distributed \cite{CL_understand,GCL_CCA}, then we have $H(\mathbf{U}^{c}|\mathbf{X},\mathbf{G}) = \sum_{i=1}^D H(\mathbf{U}^{c}|\mathbf{X},\mathbf{G}) \propto \sum_{i}^D\ln|\Sigma_{ii}|$, where $\Sigma$ is the covariance matrix of Gaussian distribution.
This indicates by minimizing the variance of features, its entropy is also minimized.
Therefore we conclude the proof.
\end{proof}
\vspace{-0.3cm}
\begin{lemma}
Minimizing $L_{unres}$ is equivalent to maximizing the entropy of \; $\mathbf{U}^{c}$, i.e., $\underset{\theta}{\min}\;L_{unres} \cong \underset{\theta}{\max}\; H ( \mathbf{U}^{c} )$.
\end{lemma}
\vspace{-0.3cm}
\begin{proof}
With the assumption that $\mathbf{U}^{c}$ obeys a Gaussian distribution, we have $\underset{\theta}{\max} \; H( \mathbf{U}^{c} ) \cong \underset{\theta}{\max}\; \ln |\Sigma _{\mathbf{U}^{c}}|$, 
where $|\Sigma _{\mathbf{U}^{c}}|$ is the determinant of the covariance matrix of the embeddings of the augmented data.
For symmetric matrix $\Sigma _{\mathbf{U}^{c}} \cong \mathbf{U}^{c\;T}\mathbf{U}^{c}$'s $D$ eigenvalues $\lambda _i$, then $\sum_{i=1}^D{\lambda _i}=\text{trace}(\Sigma _{\mathbf{U}^{c}})=\sum_{i=1}^D\sum_{j=1}^N \mathbf{U}^{c \; 2}_{ij}=N$.
We have
$\log \prod_{i=1}^D{\lambda _i} = \sum_{i=1}^D{\log \lambda _i\le D\log \frac{\sum_{i=1}^D{\lambda _i}}{D}=D\log \frac{N}{D}}$.
This means that the upper bound of $H(\mathbf{U}^{c} )$ is achieved if and only if $ \lambda_i= \frac{N}{D}$ for $\forall i$.
For $L_{unres}$, we expand it by Taylor series: $L_{unres} \propto \sum_{i}^N \sum_{j,j\ne i}^N \mathbf{U}^T_{i} \mathbf{U}_j +\frac{1}{2}(\mathbf{U}^T_{i} \mathbf{U}_j)^2 \propto \sum_{i=1}^D(\sum_{j=1}^N \mathbf{U}_{ji})^2+\frac{1}{2}\sum_{i=1}^{D}(\sum_{j=1}^N \mathbf{U}_{ji}^2)^2+\frac{1}{2}\sum_{i=1}^D\sum_{j=1,j\ne i}^D(\sum_{k=1}^N\mathbf{U}_{ki}\mathbf{U}_{kj})$. 
The $L_{unres}$ reaches the lower bound when $\sum_{j=1}^N \mathbf{U}_{ji}=0$, $\sum_{j=1}^N \mathbf{U}_{ji}^2=\frac{N}{D}$ for $\forall i$, and $\sum_{k=1}^N\mathbf{U}_{ki}\mathbf{U}_{kj}\cong 0$ for $\forall i\ne j$, i.e. the learned embeddings are uniformly distributed.
Here, $ \Sigma_{\mathbf{U}}=\frac{N}{D}\mathbf{E}$, thus the global optimum is exactly same:\;$\lambda_i=\sum_{j=1}^N \mathbf{U}_{ji}^2=\frac{N}{D}$.
Hence we have lemma 2.
\end{proof}
\vspace{-0.2cm}
The two lemmas unveil the effects of two terms in our objective. 
Combining with above analysis, referring to \cite{GCL_CCA,GCL_info}, we can further interpret Eq. \ref{eqn:loss_lu} with the task $\mathbf{T}$ of BR through following theorem:
\begin{theorem}{}  By optimizing $L_{u}$, the task-relevant information $I(\mathbf{U}^{c},\mathbf{T})$ is maximized, and the task-irrelevant information $I(\mathbf{U}^{c}|\mathbf{T})$ is minimized. Formally,
\begin{equation}
\setlength{\abovedisplayskip}{2pt}
\setlength{\belowdisplayskip}{2pt}
\begin{aligned}
\underset{\theta}{\min}\; L \approx \underset{\theta}{\max} \; I(\mathbf{U}^{c},\mathbf{T}) \text{ and } \underset{\theta}{\min} \; I(\mathbf{U}^{c}|\mathbf{T})
\end{aligned}
\label{eqn:IB}
\end{equation}
\end{theorem}
Therefore, the learned representation of user and bundle is expected to contain minimal and sufficient information about BR. Through our counterfactual loss, the noised augmented data are filtered out efficiently and the task-relevant information is kept.

\section{Experiments}

In this section, we conduct experiments on BR to evaluate the effectiveness of our method\footnote{Our code and data will be released for research purpose.}. We are aim to answer the following research questions: \quad
\textbf{RQ1:} Does our proposed counterfactual learning paradigm contribute to better performance on existing BR models? \quad 
\textbf{RQ2:} How do different designs in our paradigm influence the final performance?  \quad 
\textbf{RQ3:} Is the proposed paradigm sensitive to hyper-parameters? How do the hyper-parameters affect the effectiveness of our paradigm? \quad 

\newcommand{\tabincell}[2]{\begin{tabular}{@{}#1@{}}#2\end{tabular}}  
\begin{table}[htbp]
    \vspace{-0.3cm}
    \setlength{\abovecaptionskip}{0.0cm}
    \setlength{\belowcaptionskip}{-0.1cm}
    \caption{Statistics of datasets used in experiments.}
    \label{tab:dataset}
    \centering
    \small    
    
    \begin{tabular}{lrr}
    \toprule
    Statistic& NetEase & Youshu  \\
    \midrule
    No. of items  & 123,628 & 32,770   \\
    No. of users  & 18,528 & 8,039  \\
    No. of bundles &22,864 & 4,771 \\
    \midrule
    No. of user-bundle &302,303 &51,377 \\
    No. of user-item &1,128,065 &138,515 \\
    No. of bundle-item &1,778,838 &176,667 \\ 
    \bottomrule
    \end{tabular}
    \vspace{-0.3cm}
\end{table}

\begin{table*}[t]
    \centering
    \setlength{\abovecaptionskip}{0.0cm}
    \setlength{\belowcaptionskip}{-0.05cm}
    \small
    \caption{Experimental results (\%) of models with different training paradigms in R@\{20, 40\}, and NDCG@\{20, 40\} on two datasets. 
    For all datasets, we perform 10-times to evaluate the performance, and report the mean and standard deviations.
    The bold number indicates the improvements over the related baseline are statistically significant ($p \textless 0.01$) with paired t-tests.}
    
    \label{tab:overall}
    \begin{tabular}{p{2.6cm}<{\centering}p{1.35cm}<{\centering}p{1.35cm}<{\centering}p{1.35cm}<{\centering}p{1.35cm}<{\centering}p{0.02cm}p{1.25cm}<{\centering}p{1.25cm}<{\centering}p{1.25cm}<{\centering}p{1.25cm}<{\centering}p{0.02cm}}
    \toprule
    \multirow{2}{*}{ \bfseries Models}& \multicolumn{4}{c}{ \bfseries NetEase }& & \multicolumn{5}{c}{\bfseries Youshu } \\
    \cline{2-5}
    \cline{7-11}
    &R@20  &NDCG@20 &R@40 &NDCG@40 && R@20 &NDCG@20 &R@40 &NDCG@40 &\\
    \midrule
    BPRMF &3.512\small{$\pm 0.11$} &2.010\small{$\pm 0.10$}  &6.217\small{$\pm 0.23$} &2.734\small{$\pm 0.20$}  &&19.63\small{$\pm 0.36$} &11.17{$\pm 0.25$} &27.35{$\pm 0.29$} &13.27{$\pm 0.19$} & \\

    DAM &4.721\small{$\pm 0.21$} &2.404\small{$\pm 0.11$} &7.732\small{$\pm 0.34$} &3.210{$\pm 0.18$} &&21.02{$\pm 0.38$}  &12.05{$\pm 0.25$} &28.92{$\pm 0.47$} &14.10{$\pm 0.33$} & \\

    \midrule
    $\text{RGCN}^\dagger$ &{4.652}\small{$\pm 0.23$} &{ 2.370}\small{$\pm 0.29$} &{ 7.714}\small{$\pm 0.45$} &{ 3.170}\small{$\pm 0.31$}  &&{ 20.52}\small{$\pm 0.39$} &{ 11.52}\small{$\pm 0.40$}  &{ 28.68}\small{$\pm 0.48$}  &{ 13.67}\small{$\pm 0.39$}  & \\
    $\text{RGCN}^\dagger$+CASR &{4.782}\small{$\pm 0.25$} &{ 2.437}\small{$\pm 0.28$} &{ 7.998}\small{$\pm 0.35$} &{ 3.192}\small{$\pm 0.25$}  &&{ 20.70}\small{$\pm 0.38$} &{ 11.63}\small{$\pm 0.23$}  &{ 29.01}\small{$\pm 0.42$}  &{ 13.82}\small{$\pm 0.38$}  & \\
    \rowcolor{gray!20} {$\text{RGCN}^\dagger$+CLBR} &{\bfseries 5.210}\small{$\pm 0.21$} &{\bfseries  2.643}\small{$\pm 0.23$} &{\bfseries 8.494}\small{$\pm 0.33$} &{\bfseries 3.499}\small{$\pm 0.19$}  &&{\bfseries22.17}\small{$\pm 0.28$} &{\bfseries 11.88}\small{$\pm 0.23$}  &{\bfseries 31.05}\small{$\pm 0.35$}  &{\bfseries 14.25}\small{$\pm 0.25$}  & \\
    \midrule
    {BundleNet} &{4.776}\small{$\pm 0.41$} &{2.545}\small{$\pm 0.34$} &{8.120}\small{$\pm 0.41$} &{3.225}\small{$\pm 0.37$}  &&{22.45}\small{$\pm 0.47$} &{12.00}\small{$\pm 0.32$}  &{ 30.59}\small{$\pm 0.25$}  &{ 14.19}\small{$\pm 0.41$}  &     \\
    {BundleNet+CASR} &{4.921}\small{$\pm 0.33$} &{2.583}\small{$\pm 0.33$} &{8.320}\small{$\pm 0.39$} &{3.402}\small{$\pm 0.17$}  &&{22.45}\small{$\pm 0.48$} &{12.03}\small{$\pm 0.28$}  &{ 30.79}\small{$\pm 0.24$}  &{ 14.34}\small{$\pm 0.31$}  &     \\
    \rowcolor{gray!20} {BundleNet+CLBR}&{\bfseries 5.332}\small{$\pm 0.29$} &{\bfseries  2.711}\small{$\pm 0.24$} &{\bfseries 8.730}\small{$\pm 0.42$} &{\bfseries 3.648}\small{$\pm 0.28$}  &&{\bfseries22.53}\small{$\pm 0.60$} &{\bfseries 12.31}\small{$\pm 0.27$}  &{\bfseries 32.11}\small{$\pm 0.20$}  &{\bfseries 14.60}\small{$\pm 0.35$}  &\\
    \midrule
    BGCN &{5.760}\small{$\pm 0.48$} &{3.060}\small{$\pm 0.28$} &{9.020}\small{$\pm 0.55$} &{3.918}\small{$\pm 0.33$}  &&{22.69}\small{$\pm 0.63$} &{12.98}\small{$\pm 0.43$}  &{ 31.07}\small{$\pm 0.41$}  &{15.28}\small{$\pm 0.38$}     &\\
    BGCN+CASR &{5.911}\small{$\pm 0.36$} &{3.092}\small{$\pm 0.28$} &{9.315}\small{$\pm 0.49$} &{4.007}\small{$\pm 0.34$}  &&{22.70}\small{$\pm 0.67$} &{13.16}\small{$\pm 0.40$}  &{ 31.87}\small{$\pm 0.38$}  &{15.68}\small{$\pm 0.32$}     &\\
    \rowcolor{gray!20} {BGCN+CLBR} &{\bfseries 6.330}\small{$\pm 0.31$} &{\bfseries 3.298}\small{$\pm 0.26$} &{\bfseries 9.885}\small{$\pm 0.30$} &{\bfseries 4.255}\small{$\pm 0.22$}  &&{\bfseries 24.74}\small{$\pm 0.63$} &{\bfseries 14.10}\small{$\pm 0.32$}  &{\bfseries 34.29}\small{$\pm 0.27$}  &{\bfseries 16.73}\small{$\pm 0.32$}   &\\
    \midrule
    CrossCBR &{8.418}\small{$\pm 0.05$} &{4.565}\small{$\pm 0.03$} &{12.62}\small{$\pm 0.06$} &{5.689}\small{$\pm 0.03$}  &&{28.11}\small{$\pm 0.06$} &{ 16.68}\small{$\pm 0.05$}  &{ 37.82}\small{$\pm 0.05$}  &{ 19.37}\small{$\pm 0.04$}     &\\
    CrossCBR+CASR &{8.423}\small{$\pm 0.04$} &{4.569}\small{$\pm 0.02$} &{12.66}\small{$\pm 0.05$} &{5.699}\small{$\pm 0.03$}  &&{28.16}\small{$\pm 0.05$} &{ 16.70}\small{$\pm 0.05$}  &{ 37.85}\small{$\pm 0.06$}  &{ 19.40}\small{$\pm 0.03$}     &\\
    \rowcolor{gray!20} {CrossCBR+CLBR} &{\bfseries 8.721}\small{$\pm 0.03$} &{\bfseries 4.683}\small{$\pm 0.03$} &{\bfseries 12.85}\small{$\pm 0.06$} &{ \bfseries 5.782}\small{$\pm 0.04$}  &&{\bfseries 28.48}\small{$\pm 0.05$} &{\bfseries 16.98}\small{$\pm 0.04$}  &{\bfseries 38.30}\small{$\pm 0.03$}  &{\bfseries 19.52}\small{$\pm 0.04$}   &\\
    
    \bottomrule
    \end{tabular}
    \vspace{-0.2cm}
\end{table*}

\vspace{-0.2cm}
\subsection{Experimental Setup}
\vspace{-0.1cm}
\subsubsection{Dataset.} 
We conduct extensive experiments on two public datasets: \emph{NetEase} and \emph{Youshu}, which are widely used in the BR research \cite{DAM,BGCN}.
The statistics of all datasets after prepossessing are summarized in \autoref{tab:dataset}.

\begin{itemize}[leftmargin=*]
%
    \item \emph{NetEase} is a dataset collected by Netease from its own music platform\footnote{https://music.163.com/}. Users can bind songs into a bundle or add bundles to their favorites on this platform.
    \item \emph{Youshu} is collected by [4] from a book-review website Youshu\footnote{http://www.youshu.com/}, where every bundle is a list of books that users desired.
\end{itemize}

\vspace{-0.15cm}
\subsubsection{Baseline Models.}
To demonstrate the effectiveness of CLBR, we use following representative methods as baselines. 
\begin{itemize}[leftmargin=*]
    \item \textbf{BPRMF} \cite{rendle2012bpr} uses a Bayesian Personalized Ranking pairwise learning framework to optimize the matrix factorization model.
    \item \textbf{$\text{RGCN}^{\dagger}$} \cite{rgcn} is a classic GNN based model for heterogeneous graph, which consists of the multiple relations. 
    Here we utilize it to model the user-item-bundle graph in BR.
    \item \textbf{DAM} \cite{DAM} applies the attention mechanism multi-task learning to capture bundle-level association and collaborative signals.
    \item \textbf{BundleNet} \cite{BundleNet} constructs a similar tripartite graph to well extract bundle representations from its included items’ features.
    \item \textbf{BGCN} \cite{BGCN} is a state-of-the-art model for BR that constructs bundle-level and item-level graphs and explicitly models the user-bundle interaction, user-item interaction and bundle-item affiliation by GNN.
    \item \textbf{CrossCBR} \cite{CrossCBR} is also a state-of-the-art model which utilizes contrastive learning to model the cooperative association between bundle-level and item-level graphs to improve BR.
\end{itemize}
Among these methods, BPRMF and DAM are the representatives of traditional models without using GNN, while the rest are graph-based methods.
To verify the effectiveness of our proposed counterfactual learning paradigm, we trained the four graph-based models with different paradigms, then compare the recommendation result of every model.
It is worth noting that for a fair comparison, the graph-based models with additional side-information are not considered in our BR experiments.

As our attempt to apply counterfactual thinking in BR domain is new, there are few suitable methods for comparison. 
So we choose counterfactual method from other recommendation domain that can be adapted to BR as a reference.
Each baseline model is trained with the following counterfactual approaches:
\begin{itemize}[leftmargin=*]
    \item \textbf{CASR}\cite{CasualSR} is a counterfactual data augmentation framework for sequential recommendation, which proposed a model-orient method to choose items with larger recommendation loss and generate counterfactual sequences.
    Here we leverage this principle to sample user-bundle pairs that provide larger recommendation loss and generate counterfactual graph views for BR.
    \item \textbf{CLBR} is our proposed counterfactual learning paradigm for bundle recommendation, which can refer to section \ref{sec:model}.
\end{itemize}

\subsubsection{Evaluation Metrics.}
To evaluate the recommendation performance, we employ two widely used metrics: Recall (R@$k$) and Normalized Discounted Cumulative Gain (NDCG@$k$) following \cite{BGCN,BundleNet}, where $k=\{20,40\}$.

\vspace{-0.15cm}
\subsubsection{Implementation Details.}
We implemented the baseline models and our proposed learning paradigm based on Pytorch.
We generate four counterfactual views as the counterfactual view set $\mathcal{G}^{c}$.
The ratios $r_{ui},r_{ub},r_{bi}$ for adding/dropping user-item, user-bundle and bundle-item edges are set as hyper-parameters, which can be adjustable for different models and datasets. 
The embedding dimension is set to $64$.
The temperature parameter $\tau$ is set with a default value of $1$.
The ratios ${\alpha}^+$ and ${\alpha}^-$ are set to $0.8$ and $1.2$.
All parameters are initialized through a Gaussian distribution with a mean of 0 and a standard deviation of $0.1$.
We employ the Adam optimizer to train the models with the mini-batch size $2048$.
We conduct the grid search over hyper-parameters as follows: learning rate in $\{1e$-$5,3e$-$5,1e$-$4,3e$-$4,1e$-$3,3e$-$3\}$, learning rate decay in $\{0.01,0.05,0.1,0.5\}$, learning rate decay step in $\{2,3,4\}$, augmentation ratios in $\{0.01,0.02,0.05,$ $0.1,0.15,0.2,0.25,0.3\}$ and controlling factors $\omega_u,\omega_b$ in $\{0.01,0.05,$ $0.1,0.5,1,5,10,30\}$.
For each baseline, we adopt BPR loss and set the negative sampling rate to $1$.
The temperature factor $\tau$ in counterfactual loss is set to 1. 
We tuned all the models through our best effort. 
Through 10 times experiments with different run-time seed settings, we recorded all model's average results as well as the fluctuation range of them.


\begin{table*}[t]
    \centering
    \setlength{\abovecaptionskip}{0.0cm}
    \setlength{\belowcaptionskip}{-0.05cm}
    \caption{Experimental results (\%) of two models with different augmentation settings in Recall@20, and NDCG@20 on two datasets.
    Here we omit "@20" on the evaluation metrics for simplicity due to space constraints.
    }
    \small
    \label{tab:ablation-aug}
    \begin{tabular}{p{1.5cm}<{\centering}|p{2.2cm}<{\centering}p{0.00cm}<{\centering}|p{0.96cm}<{\centering}p{0.89cm}<{\centering}p{0.00cm}<{\centering}p{0.89cm}<{\centering}p{0.89cm}<{\centering}p{0.00cm}<{\centering}p{0.89cm}<{\centering}p{0.89cm}<{\centering}p{0.00cm}<{\centering}p{0.89cm}<{\centering}p{0.89cm}<{\centering}p{0.00cm}}
    \toprule
        \multirow{3}{*}{ \bfseries Methods}& \multirow{3}{*}{\makecell[c]{\bfseries Augmentation \\\bfseries Scheme}}& &\multicolumn{5}{c}{ \bfseries BundleNet}& & \multicolumn{5}{c}{\bfseries BGCN }\\
        \cline{4-8}\cline{10-15}
        &&&\multicolumn{2}{c}{\bfseries NetEase}& &\multicolumn{2}{c}{\bfseries Youshu}&&\multicolumn{2}{c}{\bfseries NetEase}&&\multicolumn{2}{c}{\bfseries Youshu}&\\
    \cline{4-5}
    \cline{7-8}
    \cline{10-11}
    \cline{13-15}
    && &Recall  &NDCG && Recall &NDCG &&Recall  &NDCG && Recall &NDCG\\
    \midrule
    \multirow{4}{*}{$CLBR_{SS}$} 
    &{UB only} &&{5.161} &{ 2.572}  &&{22.41} &{ 12.09}  &&{6.090} &{3.128} &&{24.24} &{13.65} &    \\ 
    &{UI only} &&{4.867} &{ 2.590}  &&{22.36} &{ 12.08}  &&{5.895} &{3.106} &&{24.12} &{13.42} &    \\ 
    &{BI only} &&{5.154} &{ 2.561}  &&{ 22.44} &{ 12.10}  &&{6.107} &{3.160} &&{24.15} &{13.68} &    \\ 
    &{UB-UI-BI} &&{\bfseries 5.212} &{\bfseries 2.682}  &&{\bfseries 22.46} &{ \bfseries 12.17}  &&{\bfseries 6.266} &{\bfseries 3.241} &&{\bfseries 24.60} &{\bfseries 13.99} &    \\ 
    \cline{1-15}
    \multirow{4}{*}{$CLBR_{M_{\theta'}S}$} 
    &{UB only} &&{5.227} &{ 2.671}  &&{ 22.50} &{ 12.25}  &&{6.264} &{3.253} &&{24.57} &{13.90} &    \\ 
    &{UI only} &&{5.145} &{ 2.582}  &&{ 22.52} &{ 12.17}  &&{5.980} &{3.175} &&{24.68} &{14.02} &    \\ 
    &{BI only} &&{5.201} &{ 2.640}  &&{ 22.48} &{ 12.21}  &&{6.184} &{3.226} &&{24.48} &{13.89} &    \\ 
    &{UB-UI-BI} &&{\bfseries 5.332} &{\bfseries 2.711}  &&{\bfseries 22.53} &{ \bfseries 12.31}  &&{\bfseries 6.330} &{\bfseries 3.298} &&{\bfseries 24.74} &{\bfseries 14.10} &    \\ 
    \bottomrule
    \end{tabular}
    \vspace{-0.2cm}
\end{table*}

\vspace{-0.2cm}
\subsection{Overall Comparison (RQ1)}
\vspace{-0.05cm}
We can obtain the following significant observations from the comparison results shown in \autoref{tab:overall}.

\noindent \textbf{Comparison of Different Baselines.} 
While training the baselines under their original methods, we can observe that graph-based models generally achieve superior results, which proves the effectiveness of graph-based models for BR task.
CrossCBR outperforms all baselines, demonstrating the effectiveness of applying graph contrastive learning in bundle recommendation.
In other graph-based models, it should be noted that BGCN  significantly perform better than BundleNet and RGCN.
We attribute the reason to BGCN's bundle-level and item-level graph modeling, which is better at differentiating users' behavioral similarity and bundles' content relatedness than BundleNet and RGCN's user-bundle-item tripartite graph, and this design is also inherited by CrossCBR.

\noindent \textbf{Observation of Paradigm Effectiveness.} 
Compared with original models, the extended models with CLBR paradigm achieve significant performance improvement on all metrics consistently, while the improvement brought by the adjusted CASR is very limited, demonstrating the superiority of our method.
For CLBR, the average improvements of each model on NetEase dataset are respectively $11.0\%$, $9.7\%$, $8.95\%$, $2.4\%$ for $\text{RGCN}^{\dagger}$, BundleNet, BGCN and CrossCBR while on Youshu dataset the improvements are $5.9\%$, $2.6\%$, $9.3\%$, $1.3\%$.
The CLBR paradigm brings the underperforming model RGCN close to SOTA graph-based models.
BGCN and CrossCBR has already obtained superior performance without data augmentation, but still can achieve vast performance improvement through our CLBR.
These results demonstrate the effectiveness and generality of our learning paradigm on graph-based models for BR.
We attribute these performance improvements to the combination of data augmentation and counterfactual constraint.
After the data augmentation of CLBR, the data used for training has the potential to better represent the integral data distribution in the counterfactual world. 
Combined with the specially designed counterfactual loss, CLBR also achieves effective noise control during the model training step, so the models could learn more accurate and BR robust embeddings for users and bundles under limited training data.
Therefore, the data sparsity problem in BR is alleviated, and the recommendations based on these embeddings show to be superior.

\begin{table}[t]
    \centering
    \setlength{\abovecaptionskip}{0.0cm}
    \setlength{\belowcaptionskip}{-0.05cm}
    \small
    \caption{Experimental results (\%) of BGCN with different counterfactual constraint designs in Recall@20 and NDCG@20 on two datasets. Here we omit "@20" on the evaluation metrics for simplicity due to space constraints.
    }
    \label{tab:ablation-loss}
    \begin{tabular}{p{3.1cm}<{\centering}p{0.8cm}<{\centering}p{0.8cm}<{\centering}p{2pt}p{0.8cm}<{\centering}p{0.8cm}<{\centering}}
    \toprule
    \multirow{2}{*}{ \bfseries Constraint Design}& \multicolumn{2}{c}{ \bfseries NetEase }& & \multicolumn{2}{c}{\bfseries Youshu } \\
    \cline{2-3}
    \cline{5-6}
    &Recall  &NDCG && Recall&NDCG\\
    \midrule
     {w/o Counterfactual loss}&{5.922} &{3.094}  &&{23.36} &{ 13.21}
    \\
    {Constant threshold} &{6.146} &{3.138} &&{24.16} &{13.59}   
    \\
    {Self-adaptive threshold } &{\bfseries 6.330} &{\bfseries 3.298}  &&{\bfseries 24.74} &{\bfseries 14.10}  
    \\
    \bottomrule
    \end{tabular}
    \vspace{-0.3cm}
\end{table}

\vspace{-0.1cm}
\subsection{Ablation Study (RQ2)}\label{sec:ablation}
\vspace{-0.05cm}

\noindent \textbf{Impact of Data Augmentation Methods.}
In this part, we compare different ways to generate the counterfactual views in CLBR, to check the impact of different data samplers and augmentation schemes.
In more detail, we substitute sampling methods and augmentation schemes of CLBR paradigm, and conduct experiments on the SOTA model BGCN, which has vast improvement using CLBR.
The method "$CLBR_{SS}$" means the augmentation comes from a stochastic data sampler, while "$CLBR_{M_{\theta'}S}$" means the augmentation with heuristic data sampler proposed in Sec. \ref{sec:phase3}, which is based on pretrained BR model $M_{\theta'}$.
For the data augmentation scheme, we compared the results of enriching the edges in BR graph individually for one relation or jointly for all relations.
All hyper-parameters are tuned to their optimal value. 

The results are presented in Table \ref{tab:ablation-aug}.
We find that both the stochastic and heuristic sampler improve the models' performance on most evaluation metrics, but the heuristic method is obviously more effective than the stochastic one. 
These results verify our conjecture about the disadvantage of stochastic mechanism in Sec. \ref{sec:phase3} for introducing too much noise in the process of generating enough counterfactual data, and prove the validness of heuristic sampler. 
Moreover, the joint augmentation for user-bundle, user-item and bundle-item relations further benefits models' performance compared to enriching each type of data separately.
We speculate that this joint augmentation can achieve a more comprehensive simulation of counterfactual distributions, which is necessary for BR.


\noindent \textbf{Impact of Counterfactual Constraint Design.}
In this part, we aim to investigate the impact of counterfactual constraint designs.
We compare the performance of BGCN model under three versions of CLBR paradigm with different counterfactual loss settings, including our proposed counterfactual loss with a self-adaptive threshold, constraint with a constant threshold, and no counterfactual loss.
As shown in Table \ref{tab:ablation-loss}, we can find that the performances of counterfactual loss versions are higher than the no counterfactual version.
By introducing the counterfactual loss, our framework can effectively filter out the remaining noise after augmentation.
When comparing the two versions with counterfactual loss, we can observe that the version with self-adaptive threshold loss outperforms the constant threshold loss version consistently.
This could be due to the fact that the self-adaptive threshold additionally considers the dynamic penalty term based on the distance between the local and global user/bundle representation distributions.

\vspace{-0.1cm}
\subsection{Hyper-parameters Study (RQ3) }\label{sec:hyper-res}
\vspace{-0.05cm}
\begin{figure}[t]
\vspace{-0.1cm}
    \centering
    \begin{subfigure}{0.45\linewidth}
        \includegraphics[width=\textwidth]{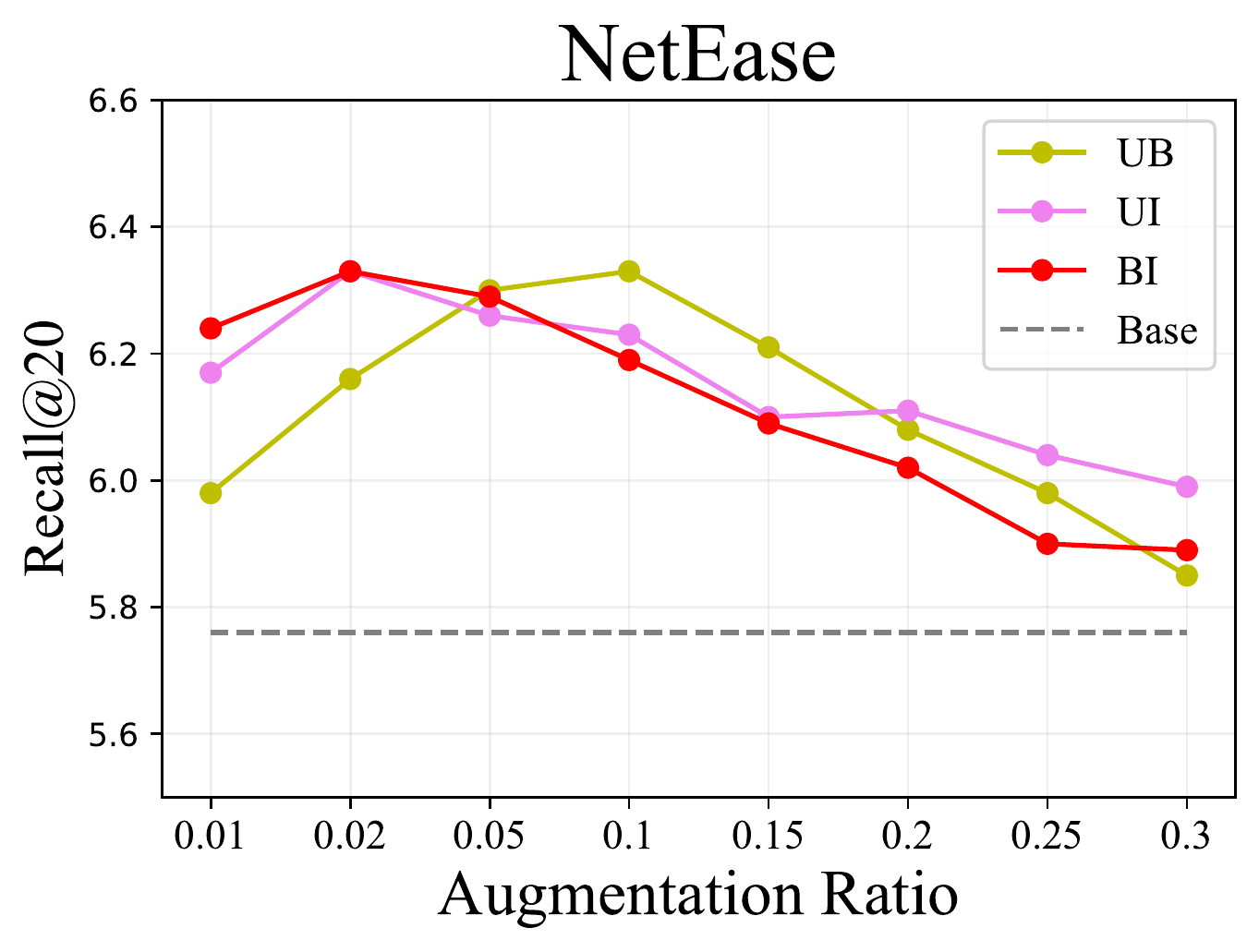}
    \end{subfigure}
    \begin{subfigure}{0.45\linewidth}
        \includegraphics[width=\textwidth]{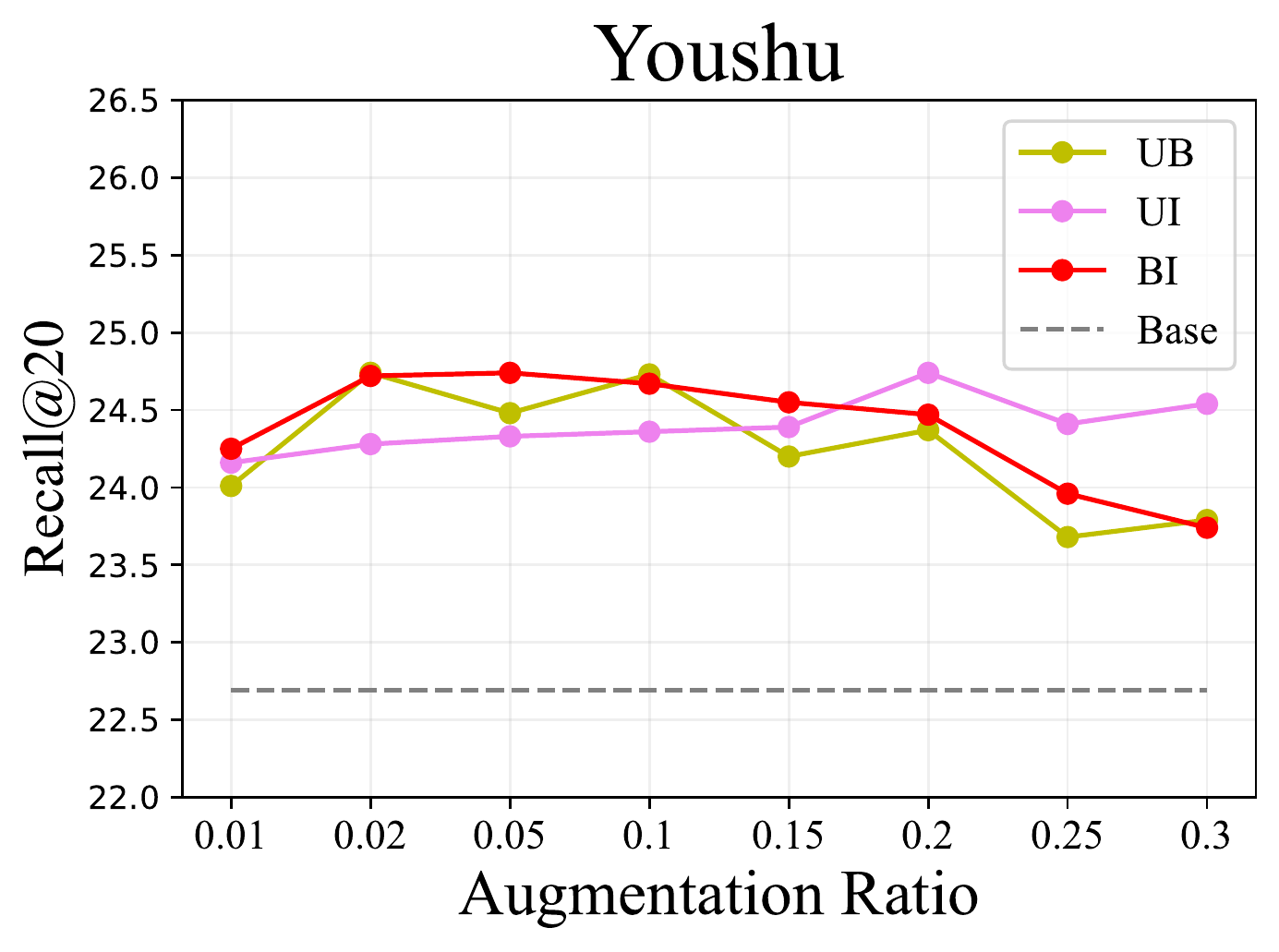}
    \end{subfigure}
    \vspace{-0.1cm}
    \caption{Influence of Augmentation ratios.}
    \label{fig:hyper-ratio}
    \vspace{-0.4cm}
\end{figure}

\begin{figure}[t]
\vspace{-0.1cm}
    \centering
    \begin{subfigure}{0.45\linewidth}
        \includegraphics[width=\textwidth]{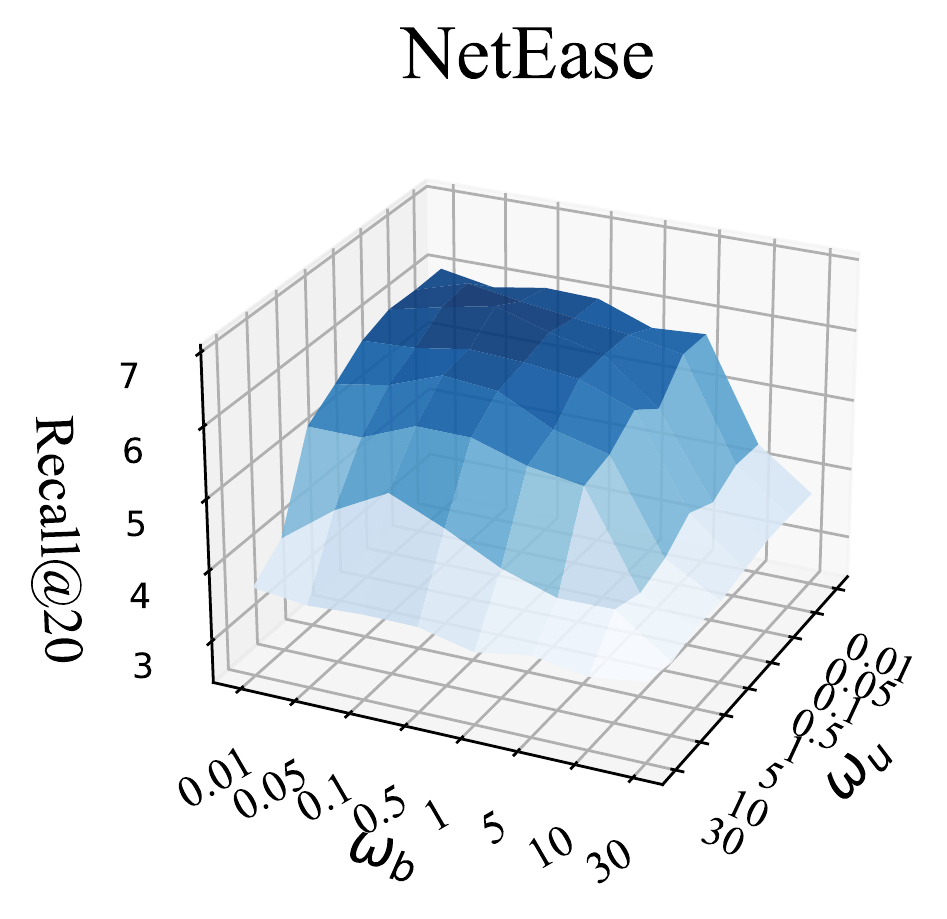}
    \end{subfigure}
    \begin{subfigure}{0.45\linewidth}
        \includegraphics[width=\textwidth]{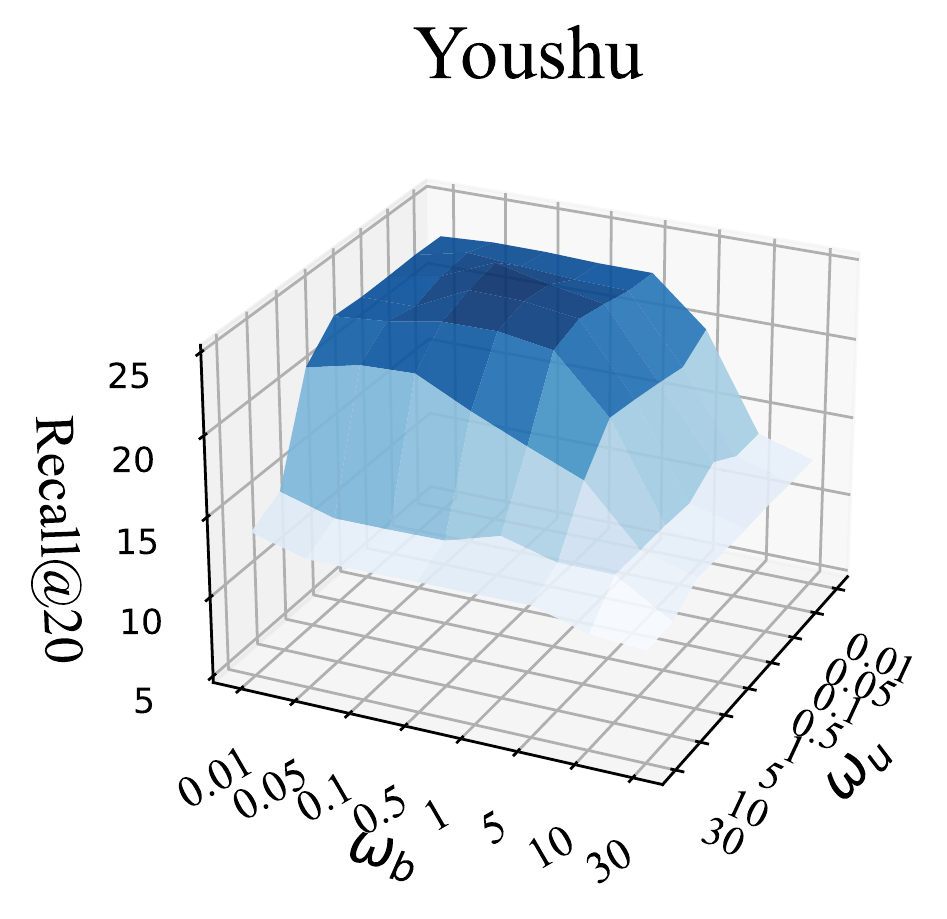}
    \end{subfigure}
    \vspace{-0.1cm}
    \caption{Influence of Controlling Factors.}
    \label{fig:hyper-omega}
    \vspace{-0.4cm}
\end{figure}

\noindent \textbf{Impact of Augmentation Ratios.}
For each augmentation ratio, we set the other two ratios and all remaining hyper-parameters to the optimal value that we've tuned in above experiments. 
Then we conduct experiments with varying the ratio from $0.01$ to $0.3$.
We present the BGCN's performance of Recall@20 on NetEase and Youshu datasets, respectively in Fig. \ref{fig:hyper-ratio}.
The results of other models are omitted here for their similarity to BGCN's and the space limitation.
The polyline "UB" shows the experiment results of different $r_{ub}$ while $r_{ui}$ and $r_{bi}$ are fixed, and "UI" and "BI" likewise denote the results of various $r_{ui}$ and $r_{bi}$.
The dotted line in figure presents the basic performance of BGCN without any augmentation.
From Fig. \ref{fig:hyper-ratio} we can discover that the performances continue to rise until reaching the optimal augmentation ratios, then drop slowly as the ratios go up.
We notice that the results remain stable when the augmentation ratios are around the optimal value, and even sub-optimal results still have considerable improvements compared to the base model.
For example, on Youshu dataset, the results stabilize above $24.65$ when tuning $r_{bi}$ in $\{0.02,0.05,0.1\}$, which are very close to the optimum.
It reveals that the effectiveness of our CLBR is not sensitive to the augmentation ratios.
In other words, we just need to adjust the augmentation ratios to a relatively loose range and then get nearly optimal results, which means the our CLBR could dramatically reduce the workload of parameter adjustment. 

\noindent \textbf{Impact of Controlling Factors.}
In Sec. \ref{sec:ablation}, we have verified that counterfactual constraint will help improve graph-based models' performance for BR.
Here we want to further explore how the weights of constraint affect the performance of CLBR.
For the controlling factors $\omega_u$ and $\omega_b$ of our proposed counterfactual loss, we fix all remaining hyper-parameters to the optimal value and conduct experiments.
We show BGCN's results using different factor combinations as 3D Surface in Figure \ref{fig:hyper-omega}.
It can be observed that on NetEase and Youshu datasets, CLBR paradigm works well when both $\omega_u$ and $\omega_u$ are in a certain range from $0.01$ to $1$.
The performance of BGCN in terms of Recall@20 is relatively stable when the controlling factors are not too large, as shown in the plateau of the figures.
In other words, the CLBR paradigm is insensitive to $\omega_u$ and $\omega_u$ in a reasonable range, which demonstrates the robustness of it.
But if we set either factor too large (e.g., $> 5$), the performance of the model is severely degraded, even much worse than the original version.
This reveals that overlarge weight makes the counterfactual constraint dominate the total loss, thus the total loss deviates too much from the BR task, which leads to a drastic performance drop.
So it is important to balance the weight of counterfactual loss by adjusting the parameters for different models and datasets.

\begin{figure}[t]
\vspace{-0.1cm}
    \centering
    \begin{subfigure}{0.98\linewidth}
        \includegraphics[width=\textwidth]{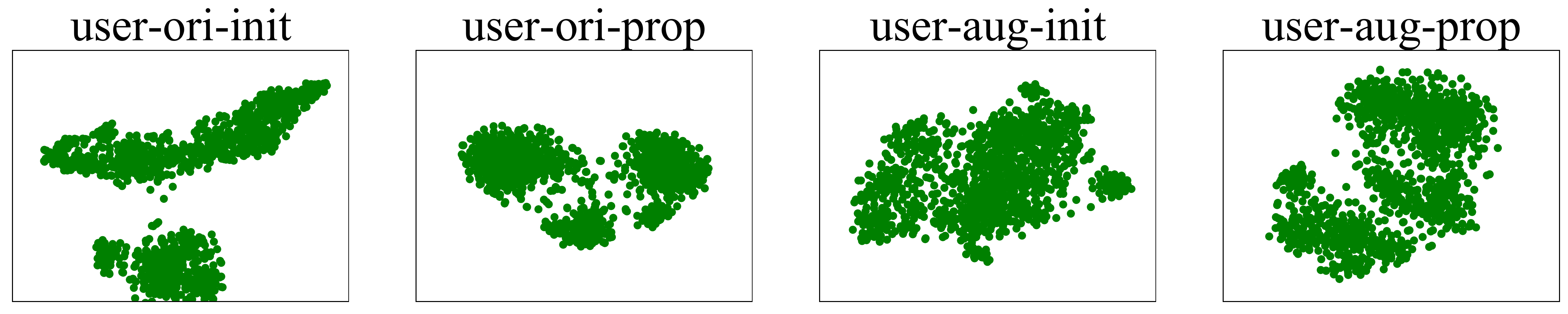}
    \end{subfigure}
    \begin{subfigure}{0.98\linewidth}
        \includegraphics[width=\textwidth]{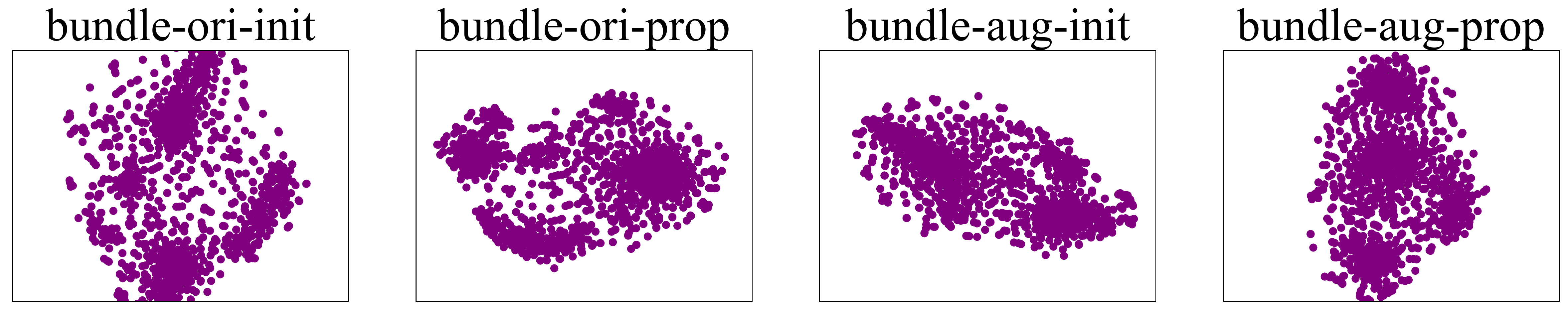}
    \end{subfigure}
    \begin{subfigure}{0.98\linewidth}
        \includegraphics[width=\textwidth]{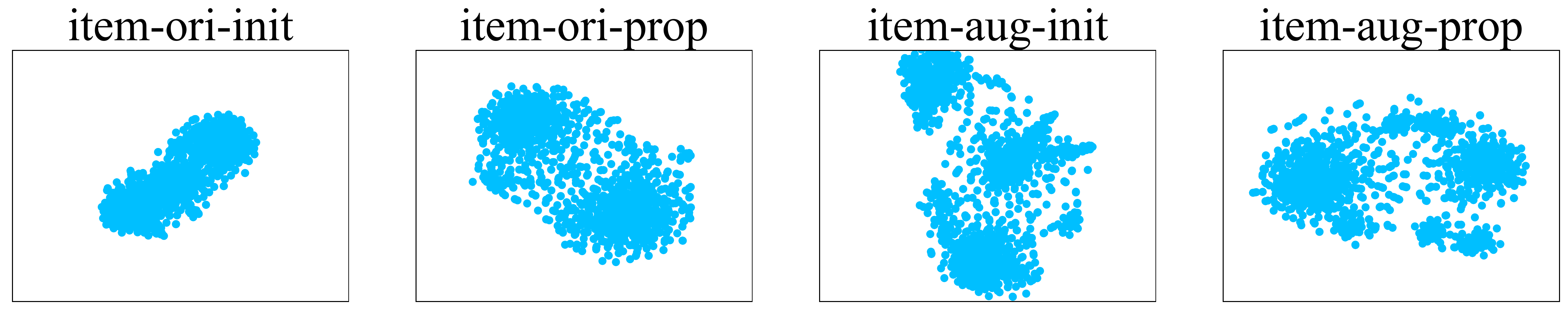}
    \end{subfigure}
    \vspace{-0.1cm}
    \caption{Embedding visualizations of users, bundles and items from Youshu dataset.}
    \label{fig:visual-analyse}
    \vspace{-0.6cm}
\end{figure}

\vspace{-0.1cm}
\subsection{Visualization and Analysis }\label{sec:case-study}
\vspace{-0.05cm}

As shown in Fig. \ref{fig:visual-analyse}, we utilize T-SNE \cite{tsne} to visualize the embedding distribution of the users, bundles and itemslearned by BGCN on Youshu dataset, to find out how does CLBR paradigm affect representation learning in BR models.
We use the "ori" to indicate the embeddings learned under original training paradigm, while "aug" means the embeddings are  learned with augmented data under CLBR paradigm.
The "init" or "prop" suffix indicates that the embeddings are obtained at initialization phase of the model or after the graph propagation. 
The nodes of user, item and bundle embeddings are colored green, purple and blue, respectively.
We get the following observations:
First, from the initialization case, compared with using original data, the embeddings initialized using augmented data evenly dispersed in representation space.
Second, the graph propagation under the counterfactual constraint of CLBR paradigm can make the embeddings distribution become multiple uniform clusters, which is aligned with our expectation for optimal embeddings.
These reveal that the augmented data has more potential to cover the counterfactual space,
and the residual noise in the augmented data is effectively suppressed under our counterfactual constraints.
Especially for users, the embeddings' distribution of original data is in an extreme disequilibrium after initialization, even graph propagation does not alleviate this situation well.
But training with augmented data effectively solves this problem. 
It again demonstrates the superiority of our proposed CLBR paradigm.
\vspace{-0.15cm}
\section{Conclusion}
\vspace{-0.1cm}
In this paper, we pay special attention to the causal view on bundle recommendation, to find a solution for the data sparsity problem and address insufficient learning of representations to improve recommendation performance.
Inspired by counterfactual thinking, we propose a novel graph-based counterfactual learning algorithm, including counterfactual data augmentation and counterfactual constraint.
We apply the CLBR paradigm to several SOTA graph-based BR models, and conduct extensive experiments on real-world datasets to verify the effectiveness of it.
As to future work, we would like to explore more interesting methods for counterfactual data augmentation. 
It's also meaningful to generalize this paradigm into more recommendation fields.

\bibliographystyle{ACM-Reference-Format}
\bibliography{ref}


\end{document}